\documentclass[universe,article,accept,pdftex,moreauthors]{Definitions/mdpi}
\usepackage[T1]{fontenc}
\usepackage{amsmath}
\usepackage{bm}
\usepackage{graphicx} 
\usepackage{multirow}
\usepackage{arydshln}

\newcommand{\earth}{\oplus}

\newcommand{\apjs}{Astrophys. J. Ser.}

\firstpage{1}
\pubvolume{10}
\issuenum{3}
\articlenumber{110}
\pubyear{2024}
\copyrightyear{2024}
\externaleditor{Academic Editors: Tamás Borkovits and Bo Ma}
\datereceived{20 October 2023}
\daterevised{16 February 2024} 
\dateaccepted{18 February 2024}
\datepublished{28 February 2024}
\hreflink{https://\linebreak doi.org/10.3390/universe10030110} 

\Title{The ``Drake Equation'' of Exomoons---A Cascade of Formation, Stability and Detection}

\TitleCitation{The ``Drake Equation'' of Exomoons---A Cascade of Formation, Stability and Detection}

\Author{Gyula M. Szab\'o $^{1,2}$\href{https://orcid.org/0000-0002-0606-7930}{\orcidicon}, Jean Schneider $^{3}$\href{https://orcid.org/0000-0002-3136-6537}{\orcidicon}, Zolt\'an Dencs $^{1,2,4}$ and Szil\'ard K\'alm\'an $^{2,5,6,}$*
\href{https://orcid.org/0000-0003-3754-7889}{\orcidicon}}

\AuthorNames{Gyula Szab\'o M., Jean Schneider, Zolt\'an Dencs and Szil\'ard K\'alm\'an }

\AuthorCitation{Szab\'o, G.M.; Schneider, J.; Dencs, Z.; K\'alm\'an, S.}

\address{%
$^{1}$ \quad Gothard Astrophysical Observatory, ELTE E{\"o}tv{\"o}s Lor\'and University,
Szent Imre h. u. 112.,\linebreak   H-9700 Szombathely
, Hungary; szgy@gothard.hu (G.M.S.); zdencs@gothard.hu (Z.D.)
\\
$^{2}$ \quad HUN-REN–ELTE Exoplanet Research Group, Szent Imre h. u. 112., H-9700 Szombathely, Hungary\\
$^{3}$ \quad Luth, UMR 1801, Paris Observatory, 
92190 Meudon, France; jean.schneider@obspm.fr \\
$^{4}$ \quad Lendület ``Momentum'' MTA--ELTE Milky Way Research Group, Szent Imre h. u. 112., \linebreak H-9700 Szombathely, Hungary 
\\
$^{5}$ \quad Konkoly Observatory, HUN-REN Research Centre for Astronomy and Earth Sciences, MTA Centre of Excellence, Konkoly Thege Mikl\'os \'ut 15--17, H-1121 Budapest,  
Hungary\\
$^{6}$ \quad Doctoral School of Physics, ELTE E{\"o}tv{\"o}s Lor\'and University, Pázmány Péter Sétány 1/A, \linebreak   H-1117 Budapest, Hungary\\}

\corres{Correspondence: xilard1@gothard.hu}

\abstract{After 25 years of the prediction of the possibility of observations, and despite the many hundreds of well-studied transiting exoplanet systems, we are still waiting for the announcement of the first confirmed exomoon.
We follow the ``cascade'' structure of the Drake equation but apply it to the chain of events leading to a successful detection of an exomoon. The scope of this paper is to reveal the structure of the problem, rather than to give a quantitative solution. We identify three important steps that can lead us to discovery. The steps are the formation, the orbital dynamics and long-term stability, and the observability of a given exomoon in a given system. This way, the question will be closely related to questions of star formation, planet formation, five possible pathways of moon formation; long-term dynamics of evolved planet systems involving stellar and planetary rotation and internal structure; and the proper evaluation of the observed data, taking the correlated noise of stellar and instrumental origin and the sampling function also into account. We highlight how a successful exomoon observation and the interpretations of the expected further measurements prove to be among the most complex and interdisciplinary questions in astrophysics.}

\keyword{planets and satellites: detection; planets and satellites: formation; techniques:\linebreak   photometric}

\begin{document}


\section{Introduction}\label{sec1}

We would like to describe briefly the astronomical picture connected with the existence and the detection of an exomoon,  which helps us in our search for satellites in distant solar systems. The fast progress of exoplanet discoveries \citep{1995Natur.378..355M,2002ApJ...568..377C,2013Icar..226.1625M,2015ApJS..217...31M,2015JATIS...1a4003R,2021ExA....51..109B} has inspired a significant interest as to whether these planets can host a detectable moon
\cite{1999A&AS..134..553S,2006A&A...450..395S,2007A&A...470..727S,2009MNRAS.392..181K}. Although no exomoon has been confirmed to date, the foreseeable observation of the exoplanets' companions will be of high interest because of a few particular reasons:

\begin{itemize}

\item{}
They reveal the internal structure of protoplanetary disks during planet formation and the secondary planetesimal formation scenarios are the protoplanetary disks.

\item{}
They provide a most precise direct observable for the dynamics in the vicinity of an exoplanet, revealing the tidal parameter, as an independent and direct constraint on the internal structure of the planet \cite{2004ApJ...610..477O,2017MNRAS.471.3019A,2024MNRAS.527.4371K}.

\item{}
They are possible abodes of exo-life, even when they are not in the habitable zone of parent stars when they are tidally heated by their parent giant planet on a wide orbit~\cite{2021PASP..133i4401D,2023AJ....165..173T}.

\item{}
A precise knowledge of its orbit around the parent planet can provide important information about the planet, such as density and mass \cite{2007A&A...470..727S}, and possibly the spin axis of the planet and its rotational oblateness \cite{2015IJAsB..14..191S}.

\end{itemize}

The lack of any undoubtedly confirmed exomoon (
a moon orbiting an exoplanet, e.g.,~\cite{1999A&AS..134..553S,2006A&A...450..395S,2007A&A...470..727S,2009MNRAS.392..181K}{)} to date appears to be a confounding conundrum of exoplanet astronomy. Detecting a moon is much more difficult than detecting a planet because of its much smaller mass and radius, while the transit of the moon also has a stochastic distribution around the central transit of the planet \cite{2009EM&P..105..385S,2019A&A...624A..95H,2022MNRAS.516.3701C}.
The power of space telescopes would enable the detection of smaller bodies than Earth in transits {(}e.g., Kepler-42 b has a radius of 0.57R$_{\rm \earth}$,~\cite{2012ApJ...747..144M}{)}, within the size range where moons surely exist---at least in our Solar System. Large moons that exist in our Solar System can regularly be formed in extrasolar systems as well \citep{2020MNRAS.492.5089M}. Despite the specific difficulties in detecting a moon, the problem of a lack of known exomoons still requires a dedicated explanation.

An already propagating naming convention is that if a companion is so large that it is just a few factors smaller than the host planet, it can also be named ``binary planets'' \cite{2007A&A...464.1133C}. There could also be moons orbiting free-floating exoplanets, as suggested by \cite{2023IJAsB..22..317R,2023MNRAS.520.5613S}.

In this paper, we review the complex processes that lead to the successful discovery of an exomoon. Following most papers in this field, we focus on a single major moon around a planet that can be detectable. We interpret this detection as the result of an event cascade in which every step must be accomplished successfully. The key steps in this sequence are:
\begin{enumerate}
\item{} Formation of a large {enough}
\endnote{In fact, the requirement of \textit{{any}} moon would suffice here as an independent scenario from \#{}3. The emphasis on the large enough size highlights the strong dependence of the occurrence on the moon size, underscoring the complex requirements even in the formation step to form and moon that is, at least presumably observable with some of the state-of-the-art instruments.} moon around a planet
\item{} Dynamical survival from its formation until our observation;
\item{} \textls[15]{Proper instrument and a verified search strategy that is decisive at the actual \linebreak system parameters.}
\end{enumerate}

All these steps are fields of research on their own. However, they are related in a very specific manner, and the discovery of the first exomoon will fully integrate these disciplines. Therefore, it is timely to examine them with respect to their interrelations and to present the structure of the problem.

In the analysis of this structure, we follow the methodology of describing the event cascade in multiplicative probability terms, just as in the case of the Drake equation \citep{1961PhT....14d..40D}. Similarly to the Drake problem, we do not want to quantitatively estimate the probability of a successful exomoon detection; we rather represent the actors in the problem, with the intention to show how the different steps have to be considered while planning exomoon-discovering strategies.

The paper is organized as follows. Section~\ref{sec2} exposes the cascade equation and explains its terms and their domains. Section~\ref{sec3} reviews key points of our current understanding of the formation, survival, and observation aspects in three subsections, while Section~\ref{sec4} discusses the results in the light of how a habitable moon can be discovered.

\section{The Cascade Equation of the Problem}\label{sec2}

To express our chances for a successful detection of a moon in an exoplanet system we sketched up the following equation:
\begin{equation}
N_{\rm moon} = N_{\rm planet} \times f_{\rm formed} \times f_{\rm survived} \times f_{\rm observable},
\label{cascade}
\end{equation}
where $N_{\rm moon}$ expresses the expectation number of our confirmed exomoon observations; $N_{\rm planet}$ is the exoplanets we observed with the appropriate technique (e.g., in transit if considering moon detection in transit light curves). The cascade terms are $f_{\rm formed}$, which denotes the fraction of moons forming in planet systems in general; $f_{\rm survived}$, which expresses the fraction of moons surviving by the time of our observation; and $f_{\rm observable}$, which is the probability of an observation of the system with our strategy---likely involving current state-of-the-art equipment.

Equation (\ref{cascade}) has a similar structure to the Drake equation: it builds up an event cascade leading to a successful discovery, consisting of conditional probabilities; and also assumes the independence of the consecutive factors. Another similarity is that the task of Equation~(\ref{cascade}) is to show the structure of the problem, in an epistemological context, and not the quantitative estimate itself directly. Equation~(\ref{cascade}) assumes an event cascade of quite independent steps that eventually lead to the detection of the moon. The moon first has to be formed, then remains in stable orbit until the observations, and then the observation has to be made. It also assumes that there is no way to observe a non-existent or an escaped moon. But the $f$ factors surely depend implicitly on several parameters of the system and the search strategy.

Therefore, the different $f$ factors are functions of various system parameters (parameters of the moon {\it {and}} 
the planet--star system), and $f_{\rm observable}$ is also a function of the observation strategy (instrument, methods, detection limits, etc.) applied to hunt the exomoons in the general case. In this way, the factors in Equation~(\ref{cascade}) depend on each other implicitly, via the various system parameters. To handle this issue, factors have to be restricted to a fixed combination of those system parameters that very significantly influence the actual value of the factors.

Let us denote the parameters of the star and the planet with $\mathcal{P}$, the parameters of the moon with $\mathcal{M}$, the parameters of the instrument and searching strategy with $\mathcal{I}$. We can now {write:} 
\begin{equation}
N_{\rm moon} = \sum _{\mathcal {P}_1}
^{\mathcal {P}_N}
\int_{\hat{\mathcal{M}}} \mathbf {f}_{\rm  \mathbf f}(\mathcal{M | P}) \ \mathbf {f}_{\rm  \mathbf s}(\mathcal{M | P}) \ \mathbf {f}_{\rm  \mathbf o}(\mathcal{M | P,I}) \ d\mathcal{M}.
\label{integral_cascade}
\end{equation}
{Here,} 
the integration goes over a set of relevant system parameters $\mathcal{M}$, covering an integration volume of a range $\hat{\mathcal{M}}$. This integral expresses how the three different fields of astrophysics and instrumentation are interrelated to each other. The enclosed integral is the estimator of an expected number of observable moons in a given system, and the summation outside the integral sums up all the known systems. Because the set of known planet--moon systems will always be finite, we can consider the individual systems by each, take the relevant system parameters into account as the conditions within the integral, and then sum up for all the known planets.

\subsection*{{The Parameter Domain} 
}

In the general case, the $\mathcal{M}$ parameter space is spanned by many variables, including the stellar mass, age, $\log g$, metallicity, abundance of refractory elements, stellar rotation rate; the six orbital elements of the planet, the planet's mass and radius, its tidal constant, its rotation period; which are known parameters (values with priors). Parameters related to the moon are also part of the domain of parameters $\mathcal{M}$, but these are eventually unknowns, and the integral has to cover them within a wide range. These include six independent orbital elements of the moon, its mass, and its size. These are 24 independent variables, and the integral will run above at least 8 variables of semi-major axes and masses of the planet and the moon, respectively, the size of the moon, the mass, age, and $\log g$ 
of the star \cite{2010MNRAS.406.2038S}; while the tidal Love number of the planet and its rotational rate are the two most important parameters above the other 8.

For calculations covering a long time span, e.g., when estimating the number of observable moons within an open cluster where even the exoplanets are not known (although we have priors for their occurrence), the complete 24-dimensional space \cite{2010MNRAS.406.2038S} has to be taken as parameters of integration in Equation~(\ref{integral_cascade}). Here, the orbital elements of the moon and the planet, and the rotation rate of all three bodies will be a function of time, described by the equations of the general three-body problem involving tidal interactions (e.g.,~\cite{2012ApJ...754...51S}), and have to be calculated as a solution of a set of differential equations for each simulation. The tidal constants of the star and the planet also evolve over time. In the most general case, we have a parameter set of 24 dimensions, and 14 of these parameters are time-dependent. Running full simulations is therefore almost impossible because of the too many dimensions and the time dependencies.

The first question is the identification of the relevant subspace of $\mathcal{M}$. Also, the relevant volume where the integrand is non-zero can be well constrained: if one of the $\mathbf{f}$ factors is $\approx$$0$ for a range of parameters, it makes any further considerations irrelevant in that parameter range. A zeroed $\mathbf f$ factor describes moons that cannot form, or/and cannot survive, or/and cannot be observed and do not offer any chance of detection. This criterion filters out most known exoplanets for exomoon discovery, which shows why we have not succeeded in a confirmed discovery of an exomoon yet. This is the task of the present contribution.

\section{The Formation Term}\label{sec3}

There are many formation scenarios known for how moons can be formed. The suggested moon formation scenarios are very diverse, and most of the multifaceted scenarios entirely depend on the star formation and planet formation processes. The most important unsolved questions related to the occurrence of observable moons today are the following:
\begin{enumerate}
\item{} What is the most effective way of moon formation? What is the maximum number of moons to form and their maximum possible size?
\item{} What are the best initial conditions that maximize the later survival chance of \linebreak the moons?
\item{} What are the stellar and planet properties that are characteristic of increased exomoon occurrence, especially the presence of large moons?

\begin{itemize}
\item{} {All} 
discussed scenarios have to be compatible with the presence of the moons in the Solar System, without assuming too specific circumstances during their formation (and hence, following the Copernican Principle).
\end{itemize}
\end{enumerate}

There are two possible origins for moons. Moons that are formed together with the planet and around it (often called \textit{{regular moons}}), and all other possible origins together (often called \textit{ {irregular moons}}).
Regular moons form around circumplanetary material before or slightly after the formation of the planet itself. The irregular moons can be formed in several other kinds of moon formation scenarios, but eventually far from the planet, and only later migrate close to the planet via gravitational capture. Moons that formed around the planet but from rings (themselves possibly being outcomes of collisions of previous giant moons) are also considered irregular moons. We review the formation terms in these two subgroups below.

\subsection{Regular Satellites}

The possible mechanisms are well reviewed by \cite{2013AsBio..13...18H,2014AsBio..14..798H}, who consider the possible formation mechanisms to lead to regular and irregular satellites and give a review of the potential habitability of these moons. Here, we compare the possible scenarios in Table \ref{formation}.

The formation of regular moons, similarly to the case of the planets, relies on \linebreak two different scenarios of core accretion and hydrodynamical disk instabilities, or even a mixture of both. Here, different pathways of formation have been recognized. During the core accretion, gravitational perturbations between planet embryos imply a series of constructive impacts up to the formation of a fully grown planet. Three different main scenarios have been proposed to describe the details of this process.

\subsubsection{Solids-Enhanced Subnebula Model}

This model assumes the formation of a subnebula \cite{2003Icar..163..198M,2009euro.book...27E}. This is composed of an optically thick inner region inside the planet's centrifugal radius (where the specific angular momentum of the collapsing giant planet gaseous envelope achieves centrifugal balance, located at $\approx$$15 R_J$ for Jupiter and $\approx$$22 R_S$ for Saturn), and the gas-to-dust ratio, around 100/1, is compatible with what is ``usually seen'' at many places in the Universe. The subnebula appears in the early stages of planet formation. The outer region of the subnebula extends to a fraction of the Roche lobe, usually $R_H/5$ is assumed. Both regions can form large moons (e.g., Io, Europa, and Ganymede are assumed to have formed in the inner region, whereas Callisto formed in the outer region). The formation is faster in the inner region. The environment is weakly turbulent in the case of both regions (e.g., \cite{2016Icar..266....1M}). At the beginning of the process, the moon is formed mostly by coagulation, and when it exceeds the radius of $\approx$$1000$~km, the forming moon develops its dynamical ``feeding zone'' and very efficiently captures the material spiraling inwards the planet. Similarly to the planets' migration, the forming moons also spiral inward in these processes, mostly due to the gas drag. The formation of a moon is quite fast and is in the order of $\approx$$10^{3}$ years for Jupiter's regular moons and $\approx$$10^{4}$ years for that of Saturn. When the circumplanetary disk disappears, the satellite system stabilizes in the short term.

\begin{table}[H]
\caption{{A summary} 
of formation scenarios proposed for exomoons. Planets show which host planets can form the moon in the scenario: G: giant; sE: super Earth; R: rocky. The Era shows whether the moon formation is synchronous with the planet formation (in situ) or happens after it (post). The relative mass shows the indicative upper limit for the $M_{moon}/M_{planet}$ ratio.}\label{formation}
\newcolumntype{C}{>{\centering\arraybackslash}X}
\begin{tabularx}{\textwidth}{cCcccC}
\toprule
& \textbf{{Process} 
} & \textbf{Planets} & \textbf{Era} & \textbf{Relative Mass} &\\
\cmidrule{1-5}
\multirow{3}{*}{\rotatebox[origin=c]{90}{Regular}} & Subnebula & G + sE & in situ & $10^{-4}$& \multirow{-2.6}{*}{\includegraphics[width=2.6cm]{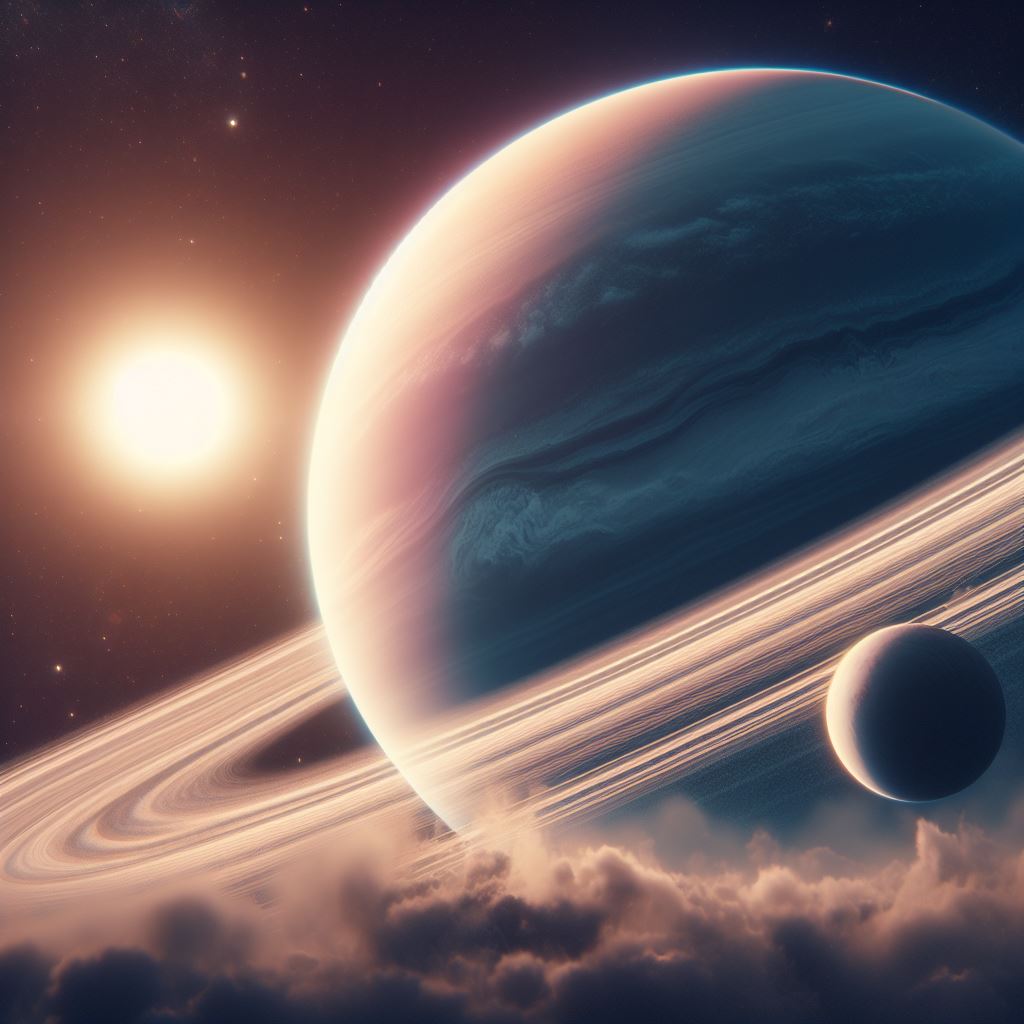}}\\
& Protosatellite disk & G + sE & in situ & $10^{-4}$&\\
& Spreading disk \vspace{2pt} & All & Post & $10^{-2}${?} 
&\\
\cdashline{1-5}
& \vspace{4pt}giant impact & R & post & $10^{-1}${??}&\\
& Tidal capture & All & post & ``no limit''& \\
\bottomrule
%
\end{tabularx}
\end{table}

Applying the description of planet formation to moons, especially by taking the ice line traps of the protoplanetary disk into account and the developments of gaps during the migration phases, {ref.} 
~\cite{2015ApJ...806..181H} discusses an elaborate scenario and concludes that ``icy moons larger than the smallest known exoplanet can form at about 15--30 Jupiter radii around super-Jovian planets'', placing these moons in the observable region. The thermodynamics of the protoplanetary envelope is a dominant factor in moon formation, and here, four sources have to be taken into account as
\begin{itemize}
\item{}viscous heating;
\item{}planetary illumination;
\item{}accretional heating of the disk;
\item{} stellar illumination.
\end{itemize}

{Ref.} 
\cite{2015A&A...578A..19H} finds that the water ice line is determining the formation of massive regular moons, and because of the lack of the water ice line, large moons cannot form regularly closer than cca. 4.5 AU to the star. But such planets, if they themselves formed behind the stellar water ice line and migrated towards the star later, can still have regular moons, so Mars-sized moons can appear even closer to the star.

One can expect that regularly formed satellites can be found around the huge number of confirmed Jupiters; thus, the subnebula model can become the most common moon formation model among the known exoplanetary systems. This model also provides examples for the formation of satellites that can be more massive than Galilei moons (e.g.,~\cite{2020MNRAS.499.1023I,2021MNRAS.504.5455C}). However, subnebula model has a significant limitation, namely, it cannot be applied to all the known satellites in the Solar System, such as moons smaller mass rocky planets, or moons that are more massive than the 10$^{-4}$ moon-to-planet mass ratio.

\subsubsection{Gas-Starved Protosatellite Disk Model}

An alternative scenario is moon formation without a subnebula, where planet satellites accrete from direct infall of gas and solids from heliocentric orbit, with the formation of a continuously replenished accretion disk around the satellites, resulting in an episodic growth of moonlets and the loss of some inward migrating moons \cite{2003Icar..163..198M,2006Natur.441..834C,2009euro.book...59C,2022A&A...667A..95O}. Formation times are on the order of $\approx$$10^6$ years within this scenario. The accretion disk can be seen as a viscous evolving protosatellite disk, characterized by peak surface densities around $100$~g/cm$^3$ \cite{2014AsBio..14..798H}.
Moons that have formed around giant planets during their growth in this environment have relative masses $\approx$$10^{-4}$ (expressing the moon mass in units of the planet's mass) \cite{2006Natur.441..834C,2010ApJ...714.1052S}.
Inside the protosatellite disk, the solid material is replenished from the surrounding protoplanetary disk. This solid material is expected to coagulate in the form of large satellites in a way similar to that of planets. Young satellites migrate inward, but some of them eventually survive sinking into the planet and are stabilized as subplanetary~companions.

A great advantage of this model is that the circumplanetary disk does not have to contain the mass required to form a Ganymede-sized relatively large moon from the beginning of the formation. Instantaneously, a less massive disk can provide moon formation. However, the disadvantage of the model is that it does not provide the formation of moons with a moon-to-planet mass ratio higher than $10^{-4}$ (just like the subnebula model). Moreover, it is necessary to solve the problem of inward-migrating moons (e.g., \cite{2021MNRAS.504.1854M}).

\subsubsection{Tidally Spreading Disk Model}

In this model, the moons are formed in a late, gas-free accretion disk that spreads out from below the Roche radius. Moons begin to form where outward spreading matter leaves the Roche radius, the region where aggregation is prevented by tidal forces from the planet \cite{2020arXiv200901881H}.  The protosatellites then continue to move outwards under the combined effects of the planetary tides and the torque from the disk. The moons grow in mass during migration and their orbit will eventually stabilize.  This pathway is considered the only way for regular moons to circle rocky planets.

In \cite{2012Sci...338.1196C}, the model assumes satellites forming \textit{{after}} the planet formation phase in a relaxed system. The model is analytical and is applicable to systems with low-mass inner moons and high-mass outer moons in the same system. They also suggested that the Earth was formed in this manner in a fast-spreading environment, which in general leads to the formation of one large satellite. (From the point of view of observations, the validity of this scenario in exoplanet systems would be ``the jackpot'', because a large and solitaire moon at a large distance from the star has by far the best chances of detecting.)

In a dynamically more relaxed environment, characterized by slow spread, ``a retinue of satellites appears with masses increasing with distance to the Roche radius, in excellent agreement with the Saturn, Uranus, and Neptune satellite systems''. This structure can explain the moon structures around Jupiter and Saturn in the Solar System.

In the detailed calculations, the exolving moon was observed to cross over three dynamically different regimes characterized by different growth scenarios. In the innermost continuous regime, the satellite is fed directly from the disk. Then, it relocates to the discrete regime and appears as a ``real'' moonlet, hydrodynamically detached from the disk. Here, the moonlets accrete the new moons appearing at the disk's edge, while they are migrating outwards into the pyramidal regime. In this outmost regime, many moons accrete altogether in hierarchical order, and at the end of this ``cannibalistic'' evolution, massive dominant moons are observed at a large distance from the planet \cite{2015ApJ...799...40H}.

However, the tidally spreading disk model provides an effective explanation for the formation of most satellite systems in the Solar System; to start the mechanism it requires a circumplanetary disk of pebbles \cite{2017AJ....153..194F}. The disk can originate from tidally captured formal planetesimals or disrupted previous-generation moonlets. Therefore, supplementary models are needed to reproduce the whole formation process.

\subsubsection{Context to Moon Occurrence}

The message of the various possible pathways of moon formation is about the large degree of complexity. For example, assumptions about the various heat sources lead to a multi-layer problem where all steps of star formation and planet formation act together, supporting a very complex environment for moon formation. Here, the time scales are important unknowns, and while the self-collapse of freshly forming planets is the main heat source for moons far out beyond the stellar water ice line, the time evolution of the planetary accretion rate will be a dominant actor in the process. The complete scenario has to be put into the context of type I-II planetary migration, a complex and not completely solved problem in itself, which is just the background of the moon formation processes.

\subsection{Irregular Satellites}

\subsubsection{Giant Impact}

Giant impacts have been the mainstream scenario suggested for the formation of the Earth's Moon \citep{2001Natur.412..708C,2004Icar..168..433C}, and it is proposed as a possible path for producing very large moons around exoplanets. In this process, a temporal circumplanetary debris disk is formed as a result of the giant impact. The disk properties influence how many moons are formed, and they proportionally scale the mass of the dominant moon in the system. Higher-mass circumplanetary debris disks tend to form a single dominant moon (such as the ``Luna'' scenario, the formation of our Moon) also called chronomoons \cite{2022MNRAS.512.1032S}, while smaller-mass disks can form multiple moons \cite{2015ApJ...799...40H}.

The possible largest mass of the moon that forms in these processes is debated in the literature. Since the Solar System example suggests a possible relative moon/planet of the order $10^{-2}$--$10^{-1}$, it is argued that the most massive moons can form after giant impacts. This high relative mass demonstrated by the Earth--Moon pair and the Pluto--Charon system may be an example as well (depending on whether Charon is a captured moon or was formed in an impact). However, \cite{2010ApJ...714L..21K} finds that roughly half of the giant impacts are followed by accretion, and they argue that a maximum mass of the debris disk, ranging in between 0.03--0.15 $M_E$, and the limited effectiveness of the accretion (only 10--50\%{} of the disk mass coagulates to moons) implies a strong mass limit around 0.07 $M_E$ and a radius around 0.15 $R_E$.

Recent simulations have shown that collisional forming of currently detectable exomoons around super-Earths is ``extremely difficult'' when accounting for only one single giant impact \cite{2020MNRAS.492.5089M}. The authors also note that it ``might be possible to form massive detectable exomoons through several mergers of smaller exomoons, formed by multiple impacts'' \cite{2020MNRAS.492.5089M}---but this field, as the case of multiple large moons in general, is very rarely discussed in the literature and is much less understood than the case of a single moon.

One of the restrictions of the canonical giant impact theory is that it does not explain the observed isotopic similarity between the composition of the Earth and the Moon. A high level of mixing between the impacter and the target body is required to resolve this problem. One or more high-energy and high-angular-momentum impacts with a fast-spinning proto-Earth can erase chemical heterogeneities by forming a synestia from the vaporized silicate material of the parent bodies \cite{2018JGRE..123..910L}. The synestia is a disk-shaped post-impact planetary structure, which rotates with the corotation limit around the targeted protoplanet since giant impacts are typical in the very last phase of planet formation. Silicates from the impacters and the mantle of the target body form a mixed continuous fluid in the synestia. Moonlets condense as the synestia cools and a moon forms by accreting the solid material (e.g., \cite{2012Sci...338.1047C}). Synestias may also serve as early tracers of moon formation; however, they have not been detected yet due to their very short life span, which takes only a few thousand years \cite{2019SciA....5.3746L}. The magnetic field can be an additional influential factor in moon formation in the case of planets that have such a relatively strong extended magnetic field like Earth. If we assume a magnetized synestia around a protoplanet, it decreases the efficiency of moon formation compared to the unmagnetized case \cite{2020ApJ...903L..15M}.

A related mode of satellite formation is the formation of moons in the rings around the planet. These rings can be seen similarly to the ``debris disks'' due to giant impacts, whereas in the case of the rings, the collision of former massive satellites has been proposed. This mode of satellite formation can be seen today in the rings of Saturn {(}see, e.g.,~\cite{2009Icar..201..191C,2012Sci...338.1052C,2012Sci...338.1196C,2020A&A...635L...8A}{)}.

\subsubsection{Tidal Capture}

Irregular satellites are often discussed as a possible way for massive warm planets to have a moon. Gravitational capture of massive moons \cite{2007ApJ...668L.167D} is often invoked, similarly to the case of irregular satellites of Uranus and Neptune \citep{2012Icar..219..737M,2018ApJ...868L..13S}.
Here, the dynamical configuration has an important influence on the outcome. Because of the need to dissipate energy during the process, capturing a single major body to a stable orbit around a planet is difficult (but can happen through chaotic orbits \cite{2003Natur.423..264A}, and subsequent tidal evolution, e.g., \cite{2011Icar..214..113N}) and is considered to occur extremely rarely. Therefore, in the baseline scenario, the capture from a former binary planet is considered. In this process, a close binary of rocky planets is broken up during a close encounter with a giant planet, resulting in one of the rocky components being rebounded to the giant planet, and the other rocky planet leaves the dynamical collision with large kinetic energy. Massive binaries with large rotational velocities and small mass ratios are most likely to undergo this process, leading to a captured moon around a giant planet \cite{2013AsBio..13..315W}. This scenario was proposed for Neptune's Triton \cite{2006Natur.441..192A}, and is considered a common source of massive moons in extrasolar systems \linebreak as well \cite{2018A&A...610A..39H}.

Tidal capturing can also be facilitated by star-giant planet-moon tidal interactions. The stellar torques can help accelerate orbit circularization, as well as spin-orbit synchronization of a loosely captured satellite in an inclined orbit \cite{2011ApJ...736L..14P}. The effectiveness of stellar torques mainly depends on the distance from the star as well as the mass of the star and the giant planet. Ref.~\cite{2011ApJ...736L..14P} pointed out that relatively close to the host star Kozai cycles may help stabilize on circular orbits of the captured moons (former terrestrial planets).


\section{The Stability Term}\label{sec4}

The stability of the exomoons is regulated by two main distance scales, the Roche radius and the Hill radius. (The Hill radius was also first described by \'Edouard Albert Roche, while, for an unambiguous naming convention, it is usually called by the name of George William Hill, who studied the dynamics of the Hill radius in detail.)
Within the Roche radius, there is no equilibrium configuration for a homogeneous satellite because it will be disrupted by tidal forces. Outside it, the satellite is held together because self-gravity exceeds the tidal forces and resists breakup \cite{1963ApJ...138.1182C}. Within the Roche radius, the orbiting material disperses and forms rings, while outside the limit, the material tends to aggregate, as has already been invoked in Section \ref{sec3}. The Roche radius depends on the radius of the planet $R_p$ and the ratio of the densities $\rho_p$ and $\rho_m$ of the planet and the moon, respectively, in the form of
\begin{equation}
R_{\mathrm {Roche}}=R_{p}\left(2{\frac {\rho_{p}}{\rho _{m}}}\right)^{\frac {1}{3}}.
\end{equation}
{This} 
can be converted to an alternative form \cite{1963ApJ...138.1182C}, expressing the orbital frequency $\Omega_{\mathrm{Roche}}$ at the Roche limit as the explicit function of $\rho_m$ as a single parameter.
\begin{equation}
{\Omega_{\mathrm{Roche}}^2 \over {\pi G \rho_m}} = 0.090068.
\end{equation}

The Hill radius is the most distant stable orbital distance of the third body in the Circular Restricted Three-Body Problem, which has a distance of \cite{1992Icar...96...43H}
\begin{equation}
R_{\mathrm{Hill} }\approx a_p(1-e_p){\sqrt[{3}]{\frac {M_p}{3(M_*+M_p)}}},
\end{equation}
where $M_*$ and $M_p$ are the mass of the star and the planet, $a_p$ and $e_p$ are the semi-major axis of the planet and the eccentricity of the planet.\endnote{These radii refer to a circular orbit approximation. If the moon has considerable eccentricity, the stability criterion ceases to be analytical at all.}
In realistic planet--moon systems, the satellites are in a complex perturbative environment of a solar system. Due to the perturbations from outside, the moon can already escape if it orbits close to the Hill radius, even if it is inside the Hill sphere. Therefore, the actual stability criterion for moons is most often rephrased with a downscaled Hill radius, in the form of
\begin{equation}
R_{\rm Roche} < a_{\rm moon, stable} < \gamma R_{\mathrm{Hill}},
\end{equation}
where the $\gamma$ factor is in the range of $1/3$ to $1/2$, based on the exact assumptions on the system (e.g., $\gamma=0.36$ for prograde moons and 0.5 for retrograde moons in \cite{1999AJ....117..621H,2002ApJ...575.1087B}, and $\gamma =0.4895$ in \cite{Domingos2006}).

Within this framework, the dynamical history of an exomoon after formation can be simply formulated: the moon, starting its life from somewhere outside of the Roche radius, is constantly receding from the planet because of the tidal interactions; and at some point, it reaches the $\gamma R_{\mathrm{Hill}}$ limit. At this point, we can practically consider that the moon is gone forever from the system. This is true if the moon initially orbits with a lower orbital frequency than the rotation frequency of the planet---but due to the basic celestial mechanics behind the formation of the moons, this assumption is very plausible.

The process is discussed in deep detail in terms of a two-body interaction in \cite{2011A&A...528A..27H}, where the differential equations are given with respect to the obliquity of both bodies, and the excentricity series goes up to the order of $e^8$. These equations have been successfully applied to calculate the tidal erosion time scale of the obliquity parameter, discussing its effect on tidal heating, the atmosphere, and habitability. Because the tidal force is scaled by the 3rd power of the semi-major axis, the tidal evolution of the orbit is very fast close to the planet, and slows down efficiently with the increasing orbital separation. This means that the most close-in, ``hot'' planets can lose their moons on a time scale of \mbox{1000--10,000 years (!}), which is way below the formation time scale of a moon---so practically, there is no space for moon formation or any stable orbit around the most close-in planets. The distant planets have a more distant Hill radius, giving the moon enough space for tidal receding and to slowly evolve the semi-major axis in an area that safely is far from the planet. These planets can keep their moons up to the time scales of 1--10 Gyrs or even more. Still, in this scenario, tidal forces heat both the planet and the moon and convert the rotational energy of the planet to orbital energy of the moon and to tidal heat. The moon is synchronized with the planet's orbital period, and its orbit will be circularized effectively.

\begin{figure}[H]
\includegraphics[width=\columnwidth]{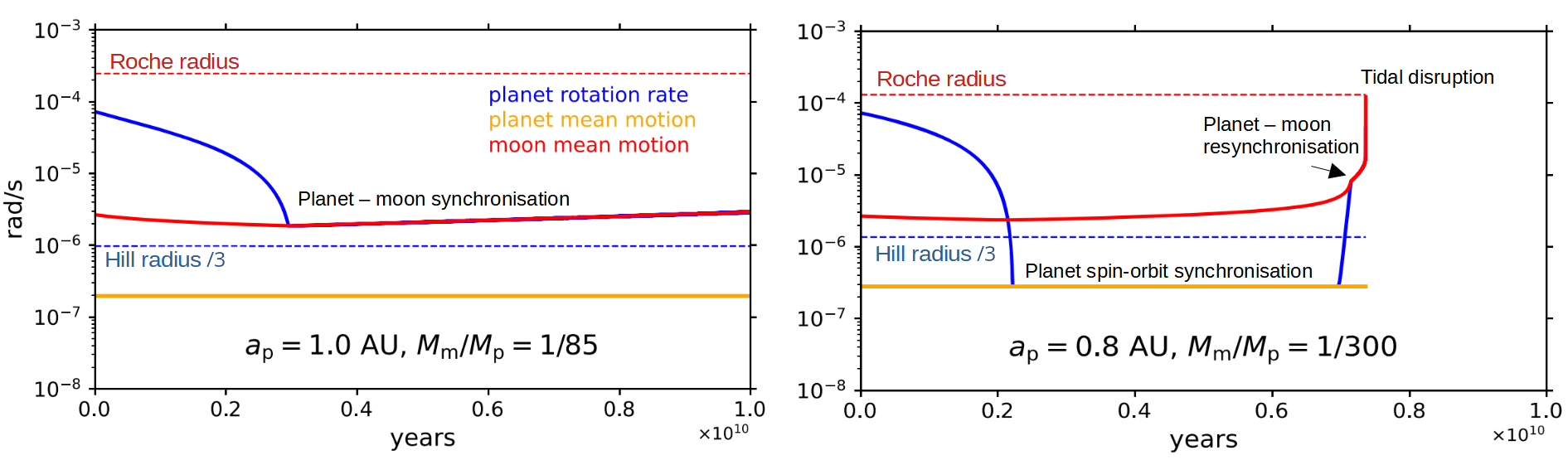}
\caption{Tidal evolution of planet--moon systems in the presence of a star. The y-axis shows units of radian/seconds, the measure of all the plotted quantities here. Yellow and red solid lines represent the planet's and moon's orbital mean motion, blue shows the rotation rate of the planet. The dashed lines show critical distances around the planets: the Roche and Hill radii, also expressed in radian/second of the mean motion on these orbits. (\textbf{Left panel}): the planet's rotation synchronizes with the moon's mean motion. (\textbf{Right panel}): first, the planet's spin and orbital mean motion synchronize. It is followed by a planet--moon resynchronization. Both simulations end with a disruption at the Roche radius, just in the left panel, this happens around 2 Gyr, beyond the shown time axis. Following \cite{2021PASP..133i4401D}.}
\label{fig:tidal}
\end{figure}

When we include the central star at the cost of neglecting the higher-order eccentricity terms and the obliquity parameter at all, we obtain a differential equation system describing the time evolution of the rotation rates (stellar and planetary rotation), and the orbital frequencies (of the planet and the moon), which also describe the evolution of the orbital elements \cite{2012ApJ...754...51S}. The behavior of the differential equation system also describes the tidal decay of the planetary rotation rate. This becomes important at the late stages of the planet--moon tidally locking (synchronization between the rotation of the planet and the mean motion of the moon); then the tidal decay of the planet's rotation (both by forces from the star and from the moon) will reduce the planet's rotation period to less than the moon's orbital period. From this point, the tidal evolution of the moon's semi-major axis reverts and the moon starts gradually approaching the planet (left panel of Figure \ref{fig:tidal}).

Needless to say, this is only possible if the moon does not escape for a long time, letting the planet's rotation slow down sufficiently. This is only possible for distant (warm and cold) planets, where the Hill radius is large, as we have already seen (Figure \ref{fig:tidal}).

At this second, ``approaching'' evolution stage of the moon, an interesting feature has also been observed in simulations. The planet initially has two options to synchronize its rotation rate: it either synchronizes to the moon's orbital rate (planet--moon locking) or to the planet's orbital rate (planetary spin--orbit locking). This latter scenario is more possible if the moon has a lower mass and a more distant orbit, so the planet--star tidal interaction outperforms the planet--moon tidal dynamics. In both cases, the moon starts approaching the planet from this point. When the moon gets close enough to the planet, the planet--moon tidal interactions will eventually supersede the star--planet interactions, when the planet's rotation rate \textit{{resynchronizes}} to the moon's orbital rate. During this stage, a huge amount of energy is dissipated and results in excessive levels of tidal heating.

From this point on, the disruption of the moon is unavoidable. The moon rapidly evolves to an orbit as close as the Roche radius, being a catastrophic end of the body and spreading away in the form of a dense ring around the planet. In some terminology, this final stage is called the ``collision with the planet'', which is a misleading term because the collision does not occur at all, whereas the moon will be entirely exterminated.

\vspace{-3pt}

The time scale of the tidal evolution is the most important factor in exhibiting an observable moon of a planet around an evolved star. In numerical experiments, simulating stochastically placed moon populations around all known exoplanets by date, ref.~\cite{2021PASP..133i4401D} concluded that in our currently known sample, only planets having an orbital period above 100--130 days have at least 50\% {} chance of keeping a moon during 1 Gyr. This also confirms that the possible targets for ``moon hunting'' observation projects must include warm or cold, ``normal'' planets; while due to the signal-to-noise criterion of observations, the host star must be very bright. The shortage of the current exoplanet catalog in these possible targets and the further perspective to discover a larger number of these will be discussed in Section \ref{discussion}.

\subsection*{{Further Parameters Contributing to the Tidal Evolution}}

In this subsection, we review the internal planet and moon parameters that have also been investigated as important actors in the tidal dynamics, while they are unaccounted for in the already discussed approximations.

The evaporation of the host planet has been suggested as an important limit for the stability of the moon, especially if the mass loss of the planet is high enough \citep{2016ApJ...833....7Y}. This mass loss leads to both shrinking Hill radius and the outward motion of moons, and as a result, global instabilities can occur that lead to the escape of moons. This scenario was found to be effective mostly for orbits with a semi-major axis $<0.1$ {AU}
. This limit is well within the typical radius belonging to the tidal stability criterion of $\approx$$0.4$~AU, and therefore observing the tidal escape limit regardless of planet evolution physics is sufficient for single-moon systems. On the other hand, {ref.} \citep{2016ApJ...833....7Y} is among the works that account for multiple moon systems instead of one major moon, and these results reveal even more complex internal dynamics in multi-moon systems.

It is crucial to emphasize that, while current research is noteworthy, it comes with certain limitations. First, these studies do not consider alterations in the rheological characteristics of the planet. This encompasses persistent dissipative features of the planetary interior, such as the K2/Q term, and the diminishing planetary radius \citep{2007ApJ...659.1661F,2013MNRAS.429..613O,2014A&A...566L...9G}. These factors play a pivotal role in the tidal evolution of a moon, particularly in the orbital transformation of larger satellites, where there is invariably an increase in the semi-major axis, increasing the risk of ejection \cite{2020MNRAS.492.3499S}, while they can reside in the system as a ploonet, a planet that is a former moon that has escaped \cite{2019MNRAS.489.2313S}. Moreover, many models often treat satellites as point masses, neglecting tidal deformations in the moons and the dissipation of orbital energy. This simplification may overlook crucial aspects, particularly if the interiors of moons exhibit fluid properties.

\section{The Term for Observability}\label{sec5}

\subsection{Transiting Exomoons}\label{sec5.1}

In the decades following the detection of the first extrasolar planet \citep[][]{1995Natur.378..355M}, the subfield of exoplanet science astronomy grew ever larger and more prominent in astronomy. This has led to the number of confirmed exoplanets exceeding 5000 in 2022\endnote{According to \url{https://exoplanet.eu}  {(accessed on 17 February 2024)} 
, the number of known exoplanets is 5633 at the time of~writing.}. Most of these planets have been discovered by the so-called transit method, enabled by the use of space observatories such as Kepler \citep{2010Sci...327..977B} and TESS \citep{2015JATIS...1a4003R}. The question of whether these planets can host a detectable moon arose almost naturally, following our knowledge of the Solar System. Several methods have been proposed for the detection of exomoons. Most of these are applicable only when (at least) the host planet is found in a configuration leading to observable transits, including the direct modeling of planet plus moon transits {(}e.g.,~\cite{2006A&A...450..395S, 2011MNRAS.416..689K}{)}, the analysis of Transit Timing Variations (of the host planet) {(}TTVs~\cite{2007A&A...470..727S, 2009MNRAS.392..181K, 2009MNRAS.396.1797K}{)}, Transit Duration Variations {(}TDVs~\cite{2009MNRAS.392..181K}{)}, Transit Depth Variations {(}also known as ``Transit Radius Variations'' or TRVs~\cite{2020A&A...638A..43R}{)} or the Rossiter--McLaughlin effect \cite{2010MNRAS.406.2038S}. Other methods, such as the detectability of radio emissions arising from the interaction between the moon and its host \cite{2014ApJ...791...25N, 2016ApJ...821...97N}, direct imaging \cite{2023A&A...675A..57K}, and the thermal signatures of the moon \cite{2021MNRAS.508.5524J} were also~proposed.

\begin{figure}[H]
\includegraphics[width = \textwidth]{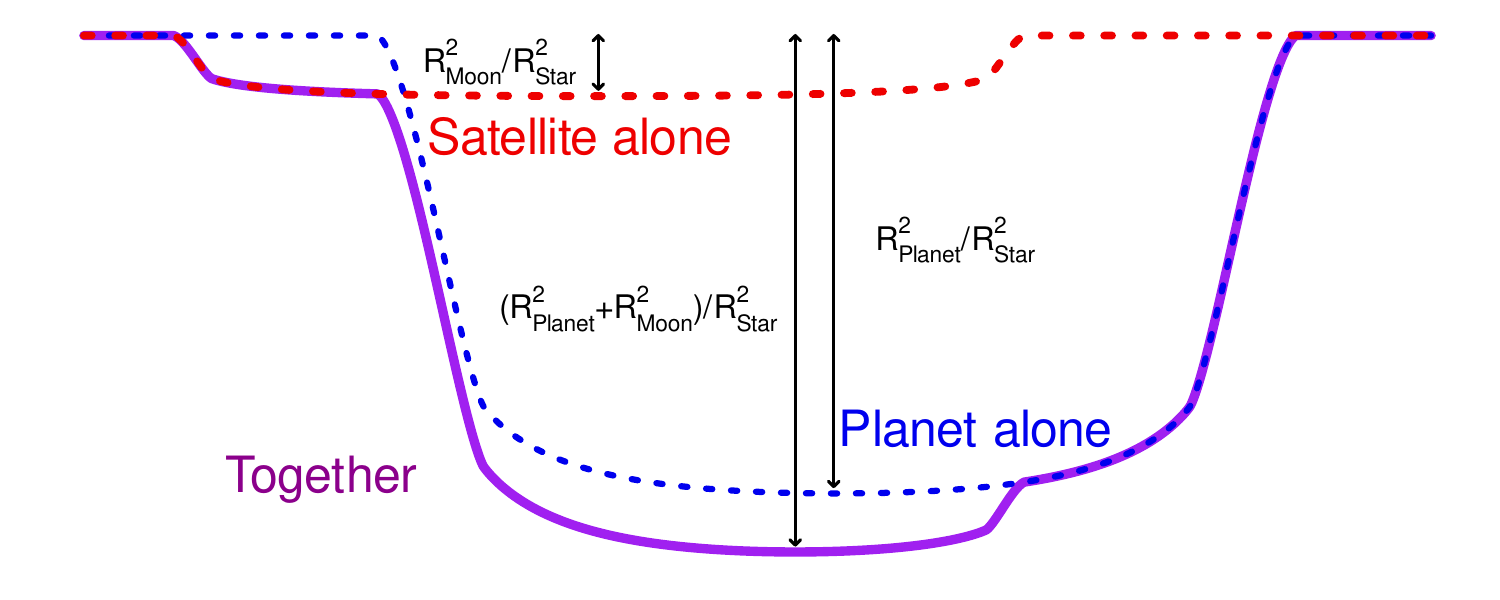}
\caption{Schematic plot showing a transit of a planet and its satellite.}
\label{fig:planet-and-moon}
\end{figure}

Inspired by the success of the Mandel--Agol model \citep{2002ApJ...580L.171M} used in the analysis of transit light curves of exoplanets, a number of models were proposed for the analysis of a planet--moon system, including an unnamed open access GUI visualization and simulator \cite{2009EM&P..105..385S}, {the} 
LUNA \citep{2011MNRAS.416..689K}, planetplanet \citep{2017ApJ...851...94L, 2019ApJ...877L..15K},
gefera \citep{2022AJ....164..111G}, Pandora \citep{2022A&A...662A..37H}, the Photodynamic Agent of the Transit and Light Curve Modeler (TLCM, \cite{2020MNRAS.496.4442C, 2024MNRAS.528L..66K}), or the analytical models of \cite{2022ApJ...936....2S, 2022AJ....164..111G} and the folding framework by \cite{2021MNRAS.507.4120K}. Generally, these are based on two fully opaque round bodies that occult a portion of the stellar disk (Figure \ref{fig:planet-and-moon}) or each other. Both the stability and detectability terms have been taken into account in the ``exomoon corridor'' analysis by \cite{2021MNRAS.506.2104T,2021MNRAS.500.1851K}, which can reveal a dynamically stable moon with transit timing analysis, even considering multiple moon systems.

A full analytical formulation of transits and phase curves is published in \cite{2022AJ....164....4L}, which can handle spherical planets and moons, and also circular starspots; taking the mutual occultations (planet--moon, planet--starspot, etc.) into account.

{Ref.} 
\cite{2014AsBio..14..798H} discusses the habitability of exomoons, in light of their formation and observability, and concludes that natural satellites in the 0.1--0.5 R$_\earth$ regime can be potentially habitable and observable. However, refs. \cite{2011IAUS..276..556S,2015PASP..127.1084S} concludes that the detection limit for an exomoon in transit is roughly identical to that of the detectable planets, which is currently the size of Mars.

\subsection{Detection Statistics}\label{sec5.2}

The detection problem belongs to the general field of {\bf {decision making}}. The crucial steps in a categorical decision process are as follows:

\begin{enumerate}
\item{} Determine a statistic from the measurements and our understanding;
\item{} Compare it to a threshold level;
\item{} \textls[15]{Evaluate the result with the possible outcomes of
Detection,  Rejection, and  Postponed~decision.}
\end{enumerate}

Detection and rejection means that the inferred signal is above or under a predefined threshold level. Firm detection means a discovery that can be published while still needing confirmation. Rejection means that the analysis showed that the data are of too low quality to meet the requirements of a reliable discovery.

A postponed decision is a third alternative when threshold levels are defined, an upper level belonging to secure detections, and a lower level belonging to secure rejections. Those cases where the test statistics rely on the two thresholds are undetermined cases: potentially interesting systems, but with no firm detection. In this case, more measurements are required.

The lower and upper thresholds are also functions of the number of data already taken, and the region of the undecided cases narrows close to zero with the increasing size of data. Setting up threshold functions can possibly be a longish procedure based on model observations and a gain function that we are optimizing for {(}see \cite{2015PASP..127.1084S} for further details{)}.

The threshold and relevance of the detections are based on the coincidence matrix that summarizes the decision statistics, as shown in Table \ref{tab:detection_statistics}. This matrix is a basis for various detection statistics. The most widespread terminology is formulated in terms of \textit{{sensitivity}} ({TPR} 
, true positive rate, what fractions of moons we find), \textit{{specificity}} ({TNR}, true negative rate, what fraction of no-moons we reject in fact), and \textit{{positive predictive value}} ({PPV}, what fraction of our detections really belong to existing moons). From these quantities, we can derive the \textit{{false alarm rate}} ({FAR}) and the \textit{{false discovery rate}} ({FDR}).
Finally, \textit{{informedness}} tells us the fraction of correct decisions above random guessing (i.e., correct decisions that are correct because of measurement). These can be formulated as in
\begin{alignat}{3}
\text{TPR}  & =  \text{TP}/(\text{TP}+\text{FN})  & & \\
\text{TNR}  & =  \text{TN}/(\text{TN} + \text{FP})  & & \\
\text{PPV}  & =  \text{TP}/(\text{TP}+\text{FP}) & &\\
\text{FAR}  & =  \text{FP}/(\text{TP}+\text{FN})  && =  1-\text{TNR} \\
\text{FDR}  & =  \text{FP}/(\text{TP}+\text{FP}) && =  1-\text{PPV}  \\
\text{INF}  & =  \text{TPR}+\text{TNR}-1  && =  \text{TPR} - \text{FAR}
\end{alignat}

\vspace{-11pt}\begin{table}[H]
\caption{Structure of the confusion matrix (contingency table) of exomoon search methods.}
\label{tab:detection_statistics}
\newcolumntype{C}{>{\centering\arraybackslash}X}
\begin{tabularx}{\textwidth}{lCC}
\toprule
&\bf {Moon} 
&\bf No Moon \\
\midrule

& {TP} 
& {FP}\\
{Found} & true positive & Fals positive\\
\midrule
& {FN} & {TN}\\
{Not found} & Fals negative & True negative\\
\bottomrule
\end{tabularx}

\end{table}

\subsubsection*{{Possible False Detections}}

At the time of writing, only a handful of exomoon candidates have been proposed. {Ref.}~\cite{2022MNRAS.513.5290D} compiled a list of potential targets for exomoon searches starting with all available planets to date and found that the top 10 planets have a more than 50\% {} chance of hosting a moon right now. Ref.~\cite{2021MNRAS.501.2378F} identified eight transiting exoplanets with TTVs in search of exomoons while claiming that the observed signal in two of these systems is unlikely to have originated from an object orbiting the planet. Ref.~\cite{2020ApJ...900L..44K} suggested that the other \linebreak six systems also produce false positive signals. A dynamical analysis by~\cite{2020ApJ...902L..20Q} also implies that KOIs-268.01, 303.01, 1888.01, 1925.01, 2728.01, and 3320.01 could not host dynamically stable exomoons. Two perhaps more well-known examples of transiting exomoon candidates have also been proposed: Kepler-1625b-i \citep{2018SciA....4.1784T} and Kepler-1708b-i~\citep{2022NatAs...6..367K}. Ref.~\cite{2019A&A...624A..95H} provide an alternative explanation for the observed signal in the former of these, suggesting that the combination of temporally correlated noise (in this case, the so-called `red noise') and Bayesian inference can yield false positive detection. Ref.~\cite{2019ApJ...877L..15K} show that by independent extraction of the HST light curves of \cite{2018SciA....4.1784T}, light curve solutions without an exomoon are preferred. Both studies \cite{2019A&A...624A..95H,2019ApJ...877L..15K} suggest that the presence of red noise on the light curves can mimic the transit of an exomoon. Negative detection of large exomoons has also recently been discussed in \citep{2023NatComm} in both the Kepler-1625 and Kepler-1708 systems. TTVs are not sufficient proof of the existence of an exomon. A TTV signal of a presumed exomoon was advertised in Kepler-1513b, the TTV signal was subsequently shown to be better represented by a second planet in the system \cite{2023MNRAS.tmp.2943Y}. As of 21 December 2023, there are only two nonretracted exomoon candidates: an exomoon at 0.13 AU of the planet MOA-2011-BLG-262L b \cite{MOA-Moon} and a volcanic exomoon around WASP-49 b \cite{Wasp-49}. Another interesting candidate is around the planet PDS-70 c, where a moon in formation may have been detected.

Given the very small number of exomoon candidates, one currently cannot predict the future percentage of false positives. For exoplanets, there are about 1\% false positives, but this statistic is based on several thousand planets.

The time-correlated noise can appear as a result of both instrumental and astrophysical noise sources, including pointing variations of the spacecraft, subpixel sensitivities, cosmic ray hits, stellar spots and oscillations, flares and microflares, granulation, etc. The presence of these effects implies the need for a modeling tool with complex noise handling capabilities, such as the wavelet formalism of \cite{2009ApJ...704...51C}, implemented in TLCM \cite{2020MNRAS.496.4442C}. Rigorous testing of this algorithm suggests that it is effective in removing red noise from light curves~\citep{2021arXiv210811822C}, producing stable parameters with correctly estimated uncertainties \citep{2022arXiv220801716K} for light curves of exoplanets. It is expected that the Photodynamic Agent update \cite{2024MNRAS.528L..66K} will also have similar success in the analysis of the transits of planet--moon systems. Furthermore, in \cite{2024MNRAS.528L..66K} we also suggest that synthetic correlated nose models can quite easily mimic the transits of an exomoon, thus drawing attention to the possibility of false positive detections.
Tidal interactions can help detection in the infrared by heating the moon, even if the companion belongs to a planet that is otherwise cold, far from the host star \citep{2021MNRAS.508.5524J}.

\subsection{Other Detection Methods Than Transit}

In the following, we are summarizing the detection methods suggested for detecting a moon, besides transit spectroscopy. While the baseline method remained the transit photometry, these alternative methods can be considered as possible ways to discover with a specific instrument (e.g., when only spectroscopy is observed with a high-performance space telescope), or independent opportunities to confirm a previously claimed detection using, e.g., transit photometry.

\subsubsection{Radial Velocity of the Parent Planet}
The extra radial velocity semiamplitude of a planet orbiting its parent star, induced by the gravitational perturbation of a moon, is given by
\begin{equation}
K = \frac{M_{moon}}{M_{pl}}\sqrt{GM_{pl} /a_{moon}}
\end{equation}

For example, for a moon of 1 Earth planet mass orbiting a planet of 0.3 Saturn mass at $4 \cdot 10^  {-2}$~AU from the planet, the radial velocity wobble of the latter will be $K = 1$~km~s$^{-1}$, detectable with an ELT class telescope.


\subsubsection{Rossitter--Mclaughlin Effect Due to a Moon}

It has been suggested that moons around transiting exoplanets may cause an observable signal in the Rossiter--McLaughlin (RM) effect. In \cite{2010MNRAS.406.2038S}, the possibility of parameter reconstruction is demonstrated from the moon's RM effect in a wide variety of planet-moon configurations. The generic parameter space has 20 dimensions with several degeneracies. The most promising parameter to reconstruct is the size of the moon, and in some cases, there is also meaningful information on its orbital period. The angles of the orbital plane of the moon can only be derived if the transit time of the moon is exactly known, e.g., from earlier transit photometry.

\subsubsection{Astrometry of the Parent Planet}

The maximum angular deviation of a planet orbiting its parent star is given by
\begin{equation}
\Delta a = 2 \frac{M_{moon}}{M_{pl}} \frac{a}{D}
\end{equation}

For a 1 Earth mass moon at 1 million km of a 0.1 Saturn mass planet at 5 pc, the maximum angular deviation is 0.3 mas. With an 8 m class space telescope, the planet image at 400 nm is 10 mas wide. A Jupiter-sized planet at 1 {au} from a solar-type star at 5 pc gives, with a 30 m diameter telescope, $N = 2500$ photons reflected from the star in a 1-h exposure in a 300 nm band. The astrometric precision of its position is then 10 mas /$\sqrt{N}$ = 0.2 mas, sufficient to detect the amplitude of the planetary wobble due to the moon revolution.

\subsubsection{Planet-Moon Mutual Events}

\begin{figure}[H]
\includegraphics[width=\columnwidth]{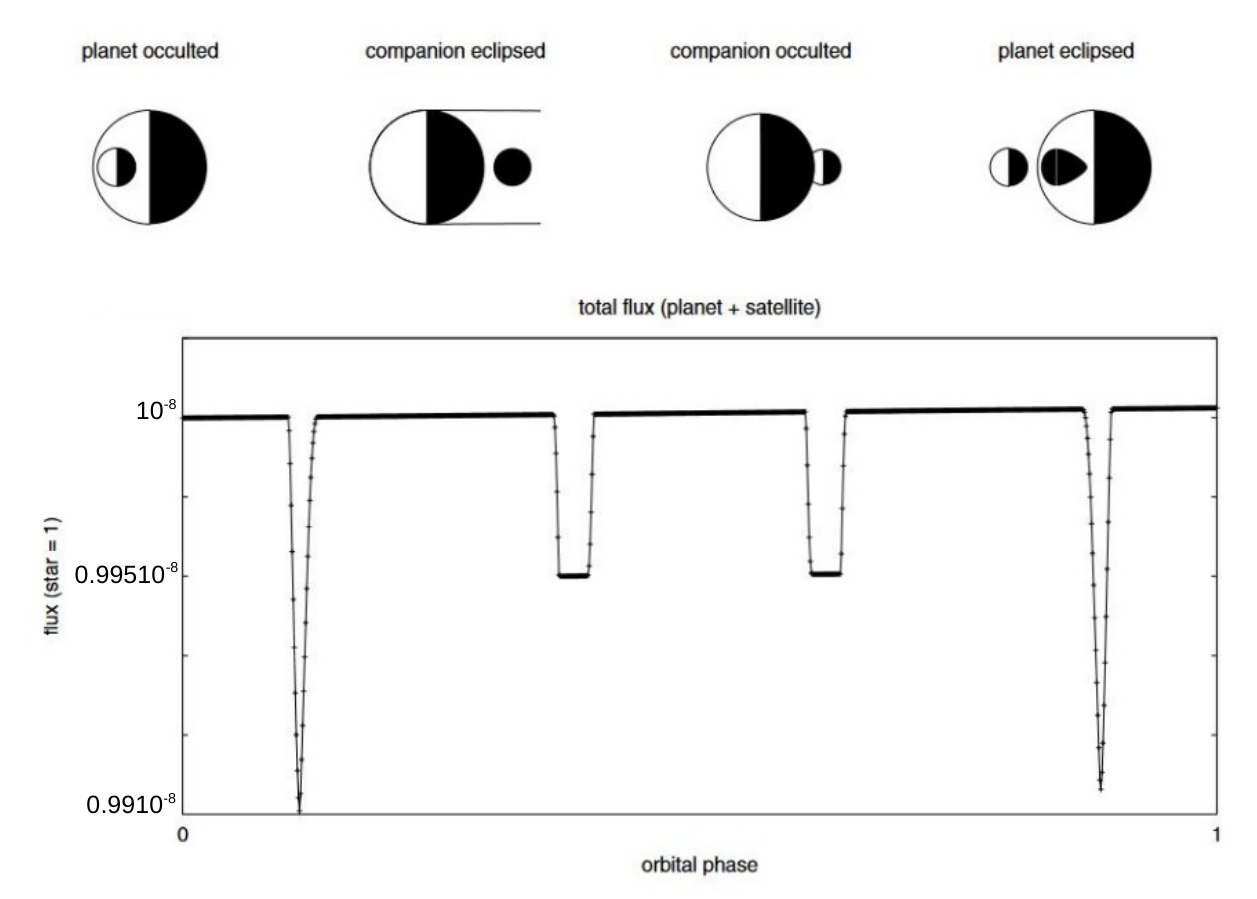}
\caption{Schematic plot showing a transit of a planet and its satellite. The vertical axis represents the total planet+moon flux as a a function of time. The dips correspond to the different mutual events shown at the top of the figure.  (following \cite{2007A&A...464.1133C}).}
\label{mutual}
\end{figure}

On average, in a direct image of the parent planet reflecting the parent starlight, the moon also reflects the starlight. However, for some parts of its orbit around the planet, it may partially mask the planet, either by transiting it or by projecting its shadow on the illuminated part of the planet. The moon is also totally eclipsed during a part of its revolution around the planet. Figure \ref{mutual} shows the resulting phase curve of the \linebreak planet + moon system reflecting the parent star.

\vspace{-12pt}

\subsubsection{Microlensing}
When a background star passes through the Einstein radius of a foreground star, we can observe an excess peak on the light curve. The Einstein radius is defined as:
\begin{equation}
R_E = \sqrt{(GM_* /c^2 ) D_S D_{LS} / D_L},
\end{equation}

When the foreground star has a planetary companion, there is a secondary peak created by the planet. When the planet itself has a moon, the background starlight curve becomes more complicated, revealing the presence of the moon, as has been suspected for OGLE-2015-BLG-1459L \cite{2018AJ....155..259H}.

\subsubsection{Direct Detection}

Extremely large ground-based telescopes will reach a $10^{-10}$ contrast, sufficient to detect an earth-sized planet. With diameters larger than 30 m they will have an angular resolution of 2.5 mas, sufficient to detect an Earth-sized moon at 0.01~AU from a planet \linebreak at 5~pc.

\subsubsection{Transit Spectroscopy of the Exomoon Atmosphere}

Transmission spectroscopy with JWST should be able to detect molecular species for the spectral retrieval of molecules in their atmosphere \cite{2010ASPC..430..139K}.

\subsubsection{Rings and Their Internal Structures}
The existence of rings is generally interpreted as evidence for former moons. Rings can be formed by the disintegration of moons that migrate below the Roche lobe \citep{2017Icar..282..195H,2023A&A...675A.174S} or by the catastrophic disruption of colliding moons \citep{2015NatGe...8..686H}. Alternative scenarios include ring formation from volcanic ejecta originating from a tidally heated satellite with active volcanism. The inner structure, especially gaps within such a ring, can also be a hint of a moon that opened the gap. Although the observation of rings will be a fairly indirect marker for an exomoon system, their discovery is technically easier than the direct detection of a moon \citep{2020A&A...635L...8A,2022A&C....4000623Z}.

\subsection{Binary Planets}
Planet--planet pairs formed in tidal capture during an instability phase of a solar system are not considered genuine ``planet moon'' systems, while--provided they \mbox{exist--would} be an easier attainable target than the smaller exomoons \cite{2024MNRAS.527.3837L}. The search for these systems is methodologically identical to the exomoon systems and will be a program, e.g., in the PLATO mission \cite{2014ExA....38..249R}.

\subsection{Combination of Detection Methods}

In addition to the above methods, one can combine some of them, leading to a richer knowledge of the planet-moon system---just as in the case of planets, where the combination of multiple detection methods leads to a better constrain of system parameters, better constrain the occurrence statistics, and even reducing the probability of false positive signals.
For example, measuring the mass of an exoplanet requires the combination of radial velocity and transit data (the latter is mostly needed to ensure a large orbital inclination).

\section{Discussion}\label{discussion}\label{sec6}

\subsection{The Structure of the Question}

The ``real'' Drake equation has a structure where the earlier terms are better known or can be better approximated, while the terms at the end are merely still unknown. Unlike this, the exomoon cascade equation has the opposite structure. The detection term at the end of the equation is the most tangible because we see the instruments for the detection, and even detailed tests can be extensively assessed. The first term, formation, is rather the most abstract in our cascade equation, because we have not seen an exomoon in formation (note here the possible exceptional case of PDS70, \cite{2021ApJ...916L...2B})   where a circumplanetary disc has been detected, possibly leading to an exomoon formation; while the orbital evolution of the moons can also be experienced via indirect observations, mostly the moon structures in our Solar System. The different methodological difficulties are seen in the timeline of the publications on the different aspects.

\begin{figure}[H]
\includegraphics[width=13cm]{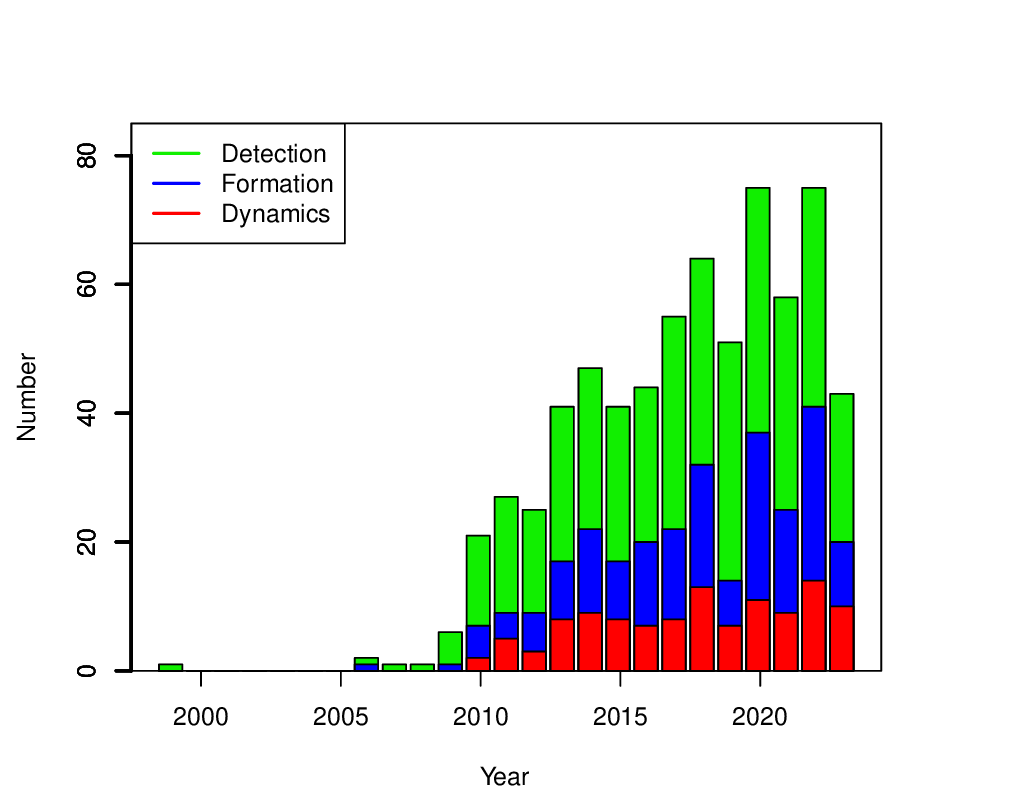}
\caption{The number of papers published each year between 1999--2023 October in the subfields of exomoon formation, dynamics, and detectability.}
\label{fig:histogram}
\end{figure}

By now, all these three fields of exomoon formation, orbital evolution, and observability have been discussed in many publications, which are summarized in Table \ref{tab1} and graphically shown in Figure \ref{fig:histogram}. The first and most widely studied question has been the observation. Here, publications cover the theory of observations (suggested methods and data analysis tools; many of these papers have been mentioned in this review); and also feasibility studies, where injections are added to simulated observations with realistic noise (white and often correlated noise parts as well), sampling, gaps, outlier points, etc., and they are finding the injection with blind solutions. The post-solution comparison of the best-fit parameters to the injection parameters returns with the reliability of the parameters, and this process can determine the detection threshold, with the aim of guaranteeing that most alerts in actual observation will represent a true positive moon detection in fact. These studies are highly driven by the development of new instruments and the always renewing research programs that dedicate more and more efforts to the exomoon search in the most appropriate approach.

\vspace{-45pt}

\begin{table}[H]
\caption{Publication and citation intensity in the subfields of exomoon formation, stability and detections on ADS\label{tab1}}
\begin{tabularx}{\textwidth}{CccCC}
\toprule
\textbf{{Subfiled} 
}	& \textbf{Number of Papers}	& \textbf{Number of Citatons} & \textbf{H-Index} &\textbf{Tori-Index}\\
\midrule
Formation & 180 & 5198 & 36 & 27.7\\
Stability & 114 & 2461 & 27 & 21.5\\
Detection & 384 & 11993 & 54 & 65.9\\
\bottomrule
\end{tabularx}
\end{table}

The moon formation studies and the dynamics are usually studied in relation to the formation of solar systems in general, in interpreting the moon formation in our Solar System, or with respect to the search for habitable worlds. It was previously discussed in this paper how extremely multilayered this is, often facing research challenges in the physical complexity of the appropriate models and practical computation limits as well. These two fields together have a similar number of publications as the observation aspects. The study of these fields started a couple of years later than the observability studies, but after revealing the actual physical processes behind a moon, the results propagated efficiently in the community, and the naive assumptions about exomoons disappeared from the data analysis work as well. (For example, no one hopes to find exomoons around hot Jupiters anymore; hence, no upper limits are derived for these, etc.)

\subsection{The Insufficiency of the Catalogs}

Due to the time scale of tidal escape that heavily depends on the orbital period \citep{2021PASP..133i4401D}, the exoplanet hosting stars can be characterized {by a} 
$P>100$ day orbital period, and the longer the period, the more chances an exomoon has to be in a stable orbit. Also, the signal-to-noise criterion is very demanding when trying to detect the exomoons because any photometric signal is inversely scaled by the second power of the radius of the exomoon (proportional to the area of the moon's disk in sky projection, much compressing the signal we wish to detect. Even so, any dynamical effect, e.g., transit timing variation, is inversely scaled by the third power of the relative radius {(}via the mass ratio; e.g., \cite{2009MNRAS.392..181K}{)}. The resulting low signal levels demand extreme precision, which is best suited by space-born instruments. Taking the actual performance into account, a reasonable limit of 10--11 magnitudes can be considered for meter-category space telescopes, and the JWST may be efficient down to 13~magnitude.

\begin{figure}[H]
\includegraphics[width=.98\columnwidth]{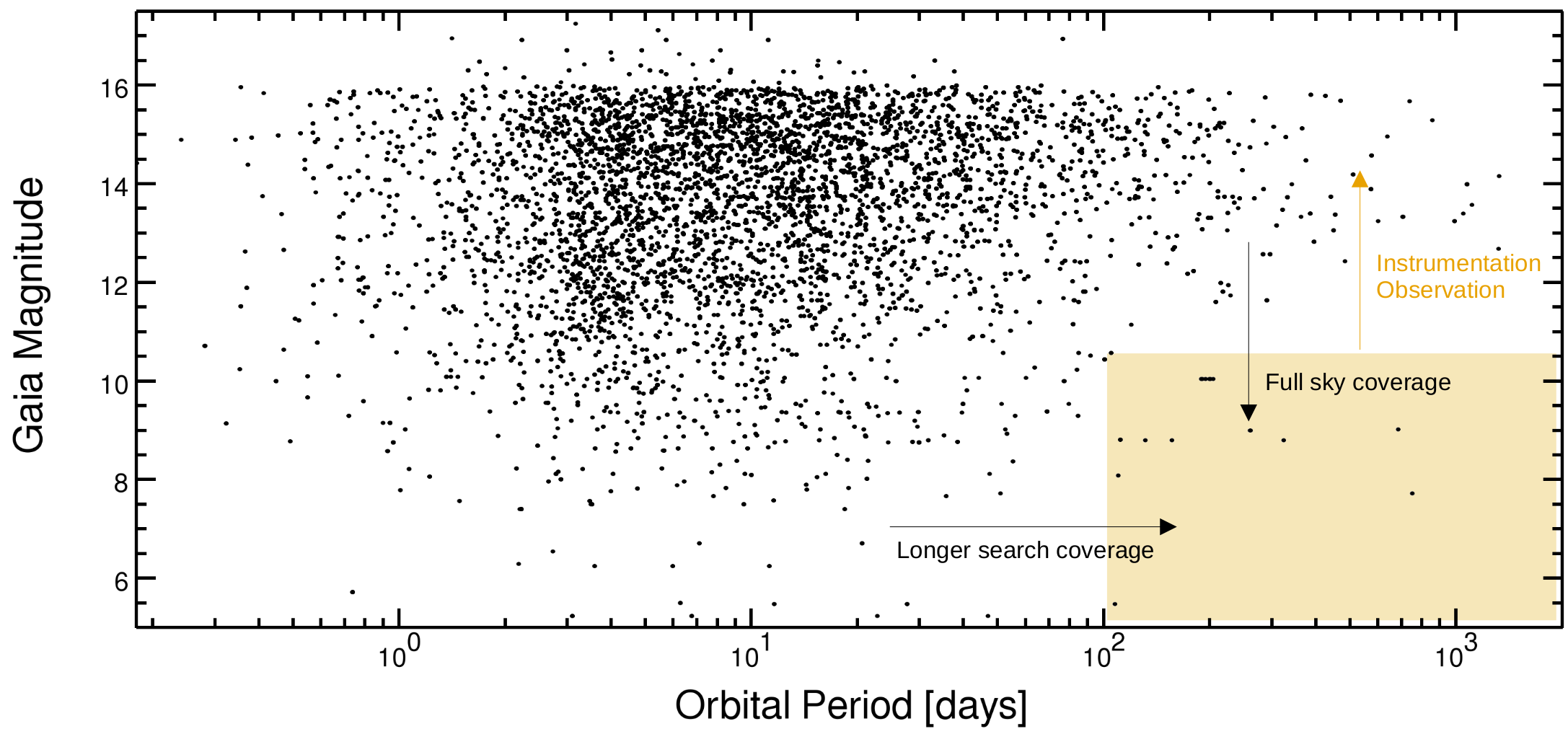}
\caption{The period---magnitude distribution of known transiting exoplanets. The colored box shows approximately our currently effective search area for moons that can astrophysically exist. Development of instrumentation and observation strategies can extend the search area upwards while increasing time coverage and sky coverage of the search programs give more potential targets into the search area.}
\label{region}
\end{figure}

Taking these constraints in mind, an ``exomoon search region'' can be defined as $P>100$~d, $G<11$~mag. Despite the more than 5500 confirmed exoplanets so far, we still have an insufficient number of planets in the exmoon search region. Figure \ref{region} shows the distribution of the confirmed exoplanets in the period--magnitude space, having the search region overplotted. There are a few planets within the region, which is summarized \linebreak in Table~\ref{catalogs}.

\begin{table}[H]
\caption{{The} 
known systems with planets over 100 day period and host stars brighter than \mbox{$G = 11$ mag}. Remarks: {$^{\dag}$}
: the approximate periods of the three planets; {$^{\ddag}$}: the solution is not~unique.}
\label{catalogs}
\newcolumntype{C}{>{\centering\arraybackslash}X}
\begin{tabularx}{\textwidth}{LCCCC}
\toprule
\textbf{Host Star} &\textbf{\boldmath{ No. Planets $P > 100$ d }}&\textbf{ Period [d]} & \textbf{{G} mag} & \textbf{Discovery} \\
\midrule
HD 136352 & 1 & 107.245 & 5.485 & \cite{2021NatAs...5..775D} (2021)\\
GJ 414 A & 1 & 749.83 & 7.720 & \cite{2021AJ....161...86D} (2021)\\
HD 114082 & 1 & 109.75 & 8.094 & \cite{2022AA...667L..14Z} (2022)\\
HIP 41378 & 3 &  278, 369, 542 $^{\dag}$ & 8.810 & \cite{2016ApJ...827L..10V} (2016)\\
HD 80606 & 1 & 111.436 &  8.820 & \cite{2009AA...502..695P} (2009)\\
TOI-2180 b & 1 & 260.79 & 9.011 & \cite{2022AJ....163...61D} (2022)\\
TIC 172900988 A & 1 & 188--204 $^{\ddag}$ & 10.048 & \cite{2021AJ....162..234K} (2021)\\
Kepler-126 & 1 & 100.283 & 10.454 & \cite{2014ApJ...784...45R} (2014)\\
TOI-199 & 1 &  104.854 & 10.578 & \cite{2023arXiv230914915H} (2023)\\
\bottomrule
\end{tabularx}

\end{table}

Somewhat surprisingly, the number of promising targets in our search area has increased very significantly in the past few years, and 6 out of 9 systems were discovered in 2021 or later. This rapidly increasing rate of discoveries can be devoted to the long observation time basis for planet search (e.g., repeated TESS visits) and the evolution of the observation strategies (e.g., follow-up long-period TESS planet candidates to determine the accurate period, {(}e.g.,~\cite{2021NatAs...5..775D,2023A&A...674A..43U,2023A&A...674A..44G,2023MNRAS.523.3069O}{)}, which help the precise characterization of long-period transiting planets. The chances of obtaining more and more appropriate candidates within the exomoon search region are increased due to the longer time coverage of the observations and also the better resolution of the full sky with appropriately long and precise observations {(}
as indicated by the arrows in Figure \ref{region}.

Most of the current and future space observatories have an exomon program. The JWST is capable of detecting super-Titans around isolated planetary-mass objects~\mbox{\cite{2021ApJ...918L..25L,2022BAAS...54e.183L}}, and when the homogeneous surface temperature distribution is taken into account, Titan category exomoons can also be considered \cite{2021MNRAS.508.5524J}. CHEOPS \cite{2021ExA....51..109B} has also attempted to detect large exomoons around long-period planets the large exomoons \cite{2023A&A...671A.154E}, PLATO will be searching for exomoons from transit photometry \cite{2014ExA....38..249R}, while Ariel will also have a program to detect exomoons and rings using photometers in the optical and synthetic spectrophotometry in the infrared~\cite{2022ExA....53..607S}.

\subsection{Open Questions}

Exomoon studies have seen intense development in the past years, and by now the field has a long frontline that progresses into the unknown. The open questions addressed here are by no means considered to be comprehensive; rather, they are selected aspects of the ``exomoon chase'' efforts, which the authors consider to be especially important, and the role of this list is to propose further research directions for the interested expert. The structure follows the formation--dynamics--observation arrangement.

\begin{enumerate}
\item{Formation}

A central question concerning formation is \textit{{what kind of a moon}} we are looking for and how large exomoons we can hope to observe. Once we decide our threshold, we have to answer whether there is an upper limit for a moon in mass or size. What is the best target for which we are hunting?

What is the occurrence of exomoons, and especially, what is the occurrence of these ``large enough'' moons around different types of planets, and what is the size dependence of the occurrence? Does the exomoon occurrence in general have some spatial dependence in the well-observable parts of the Galaxy?

Are there observable diagnostics or proxy parameters belonging to an exomoon? Are there special stellar and/or planetary parameters that, at least in combination, can serve as a red light for a large exomoon? Understanding these aspects is conceptual for observation work.

In addition, the validity of different formation pathways for moons is a debated question. Which of the suggested processes is acting in real systems and what is the parameter preference of the different modes? There are still many aspects to understand about the possible formation methods of the Solar System moons as well.

\item{Dynamics}

How well do the current equations describe the orbital evolution of the moons? Tidal dynamics is a very complex process in itself, whereas in real systems this will be embedded in the N-body problem of a whole solar system {(}e.g.,~\cite{2023AJ....166..208H}{)}. Are we still missing some key aspects that can rearrange the entire picture of our current understanding here?

Tidal heating itself also has a wide horizon in habitability studies and the search for a life-supporting environment. Here, the key question is the actual eccentricity of the systems, the initial distribution of the eccentricity, and the time scale and precision of the circularization processes.

\item{Observations}

From a technical point of view, high-level signal reconstruction is required. Here, the adaptive handling of possible known sources of instrumental effects, such as data from telemetry and imaging and the less predictable sources like stellar noise, is a central question that can help get rid of the several (very probably) false positive detections that have already been discussed. (Here, we can remark that at the early stages of exoplanet discovery, there were also a meaningful number of false positives, representing the inherent characteristic of very difficult observations.)

An open question for observations is that, after the first discovered exomoons, how can we learn something about the internal structure of exomoons. Atmospheric spectroscopy and thermal photometry are two suggested pathways for the future to reveal the reality of these still unknown worlds.

One of the most exciting possibilities is the imaging of exomoons. This is a reasonable hope with future ground-based (like the E-ELT) or lunar very big instruments. For instance,  (1) a 10 $\times$ 10 pixel image of an icy exomoon with a lunar hypertelescope \cite{2021ExA....51.1003L} will reveal plumes, like for Europa (2) if the surface of the exomoon is inhomogeneous, multiple subsequent images will tell if the moon is co-rotating with its parent planet (like our Moon); that will give another information of the moon internal structure.

\end{enumerate}

\subsection{Habitability and Civilization}
Although this paper is mainly descriptively pointing toward the steps that lead to the observation of an exomoon, during discussions we can speculate on how the \textit{actual} Drake equation looks like if we introduce terms that allow life on exomoons as well, besides the exoplanets. At first glance, we could imagine that life emerging on planets and on moons can be additive, represented by a second term added to the Drake equation, allowing life to emerge on moons. In this formalism, following the usual notations to express the number of communicating civilizations in the Milky Way as $N$, we could naively write
\begin{equation}
{N=R_{*}\cdot
\left(
f_{\mathrm {f} }^\mathrm{p}\cdot n_{\mathrm {e} }^\mathrm{p}\cdot f_{\mathrm {l} }^\mathrm{p}\cdot f_{\mathrm {i} }^\mathrm{p}\cdot f_{\mathrm {c} }^\mathrm{p}
+
f_{\mathrm {f} }^\mathrm{m}\cdot n_{\mathrm {e} }^\mathrm{m}\cdot f_{\mathrm {l} }^\mathrm{m}\cdot f_{\mathrm {i} }^\mathrm{m}\cdot f_{\mathrm {c} }^\mathrm{m}
\right)
\cdot L},
\label{Eq:original}
\end{equation}
where
and $R_{*}$ is the average rate of star formation in our Galaxy;
$f_\mathrm {f}^\mathrm{p}$ is the fraction of those stars that form planets,
$n_\mathrm {e}^\mathrm{p}$ is the average number of planets that can potentially support life per star that has planets,
$f_\mathrm {l}^\mathrm{p}$ is the fraction of planets that could support life that actually develops life at some point,
$f_\mathrm {i}^\mathrm{p}$ is the fraction of planets with life that go on to develop intelligent life (civilizations),
$f_\mathrm {c}^\mathrm{p}$ is the fraction of civilizations that develop a technology that releases detectable signs of their existence into space, and
$L$ is the length of time for which such civilizations release detectable signals into space. In the second term, the $f$ and $n$ factors represent the same coefficients, but for the moons.

This formalism is naive mostly because it does not assume that the formation of an intelligent civilization does require time, and if we take the single observable civilization into account, we have to estimate the time scale to be in the order of 4--5 billion years. Only F and later spectral-type stars meet this criterion. However, the initial mass function of star formation describes that most stars that actually form are low mass enough to have a longer Main Sequence life than this time scale. In this way, although the time scale for the formation of intelligent life is not included in the Drake equation, this does not mean a severe constraint {{of planes}}.

In the case of moons, the situation is rather different, mostly due to the tidal dynamics of moons. We know that {most of the} moons that could have formed around the already known planets would have escaped or been disrupted by now and would not support life at all. Also, there is the possibility for life on moons rogue planets that are either so far from their host star that the incident stellar flux does not influence their thermodynamics, or that have completely escaped from a solar system. If they have a moon on an eccentric orbit enough, tidal heating itself can support the habitability of the moons, and these can be taken into account as potential places of life. This could have been expressed as another additive factor in Equation~(\ref{Eq:original}). A third point is that although Equation~(\ref{Eq:original}) simply expresses $f_\mathrm {i}^\mathrm{p}$ as a factor independent of $f_\mathrm {f}^\mathrm{m}$, they are actually connected. We do know the important role our Moon has played in developing our cosmographically informed civilization and our interest in finding alien civilizations by communication \citep{1979ossi.book.....J}. Therefore, the presence of a moon may be considered as a strong support, if not a must for a nonzero $f_\mathrm{i}$ in the Drake~equation.

This is a merely philosophical aspect, due to the large number of unknown factors, which would merit a dedicated review of its own. But in essence, we can conclude that moons that are present for a long enough time in solar systems can substantially contribute to the development of comsographically informed civilizations.

\section{Concluding Remarks}\label{sec7}

We reviewed the processes leading to exomoon detection along the steps described in Equations (\ref{cascade}) and (\ref{integral_cascade}). In qualitative terms, we may recognize the role of many different cut-offs in the parameter space of the three different factors in the cascade. For a successful observation, a (1) very large moon, (2) very far from the planet, which (3) has a short orbital period, would be the best target, simply to support frequently observable transits where the moon component is evidently present with a high signal-to-noise ratio. This system evidently does not exist because of its dynamical instability on a very short time scale.

Unlike the ``original'' Drake equation where the later factors are less known, in our cascade the factors stand in the opposite order. The term for observability can be fairly well quantified by means of synthetic observations and detection evaluation. The stability term is also deterministic in its explicit form, but it also relies on the tidal dissipation coefficients and subtle fine-tuning of parameters (e.g., eccentricity) that we do not know precisely. In this sense, the stability is less quantifiable than the observation term. However, it generally supports moons far from the planet (which is very favorable from the observational point of view) and planets on long-period orbits (which favors longer time between transits, slowing down the discovery process).

The formation term itself has very complicated physics in the background, connecting many aspects of planet formation as the outer environment of moon formation. This term is very difficult (if not impossible) to quantify in general, but evidently, it includes very steep cut-offs towards very large moon size and moon mass. The trade-offs select the character of the discoverable moons in parameter space: The sweet spot is possibly the quest for smaller (but still observable) moons around long-period planets. This is our best chance, even if the catalogs are short here (although such systems are just being discovered in an increasing number) and the size limit for a positive detection is still quite large, at least as large as Ganymede.

In 2024, the study of exomoons \cite{1999A&AS..134..553S} is seeing its 25th anniversary. During the past quarter of a century, the true astrophysical and observational complexity of the task, and the actual impediments on the way to the first confirmed detection. The following years are very promising in this sense because we are just in the position of discovering a favorable amount of long-period planets around enough bright stars. In \cite{2019arXiv191112114H}, it is suggested that $\sim$100,000 exoplanets may be discovered by 2050. Therefore, there is reasonable hope that we will have thousands of long-period planets that can host a moon physically. If the detection rate of exomoons is on the order of a few percent, we can expect dozens or maybe hundreds of confirmed exomoons by the 50th anniversary of the field.

\vspace{6pt}


\authorcontributions{Conceptualisation, writing, editing and publishing: G.M.S. General consulting, {Section} 
\ref{sec5.2} 
, Section \ref{sec1} paragraph 1: J.S. Structural idea and literature review for Sections \ref{sec3} and \ref{sec4}: Z.D. Full elaboration of Section \ref{sec5.1}, correspondance and submission: S.K. {All authors have read and agreed to the published version of the manuscript.} 
}

\funding{S.K. and G.M.S. thank the PRODEX Experiment Agreement No. 4000137122 between the ELTE E\"otv\"os Lor\'and University and the European Space Agency (ESA-D/SCI-LE-2021-0025). Project no. C1746651 has been implemented with the support provided by the Ministry of Culture and Innovation of Hungary from the National Research, Development and Innovation Fund, financed under the NVKDP-2021 funding scheme. Project no. 147362 has been implemented with the support provided by the Ministry of Culture and Innovation of Hungary from the National Research, Development and Innovation Fund, financed under the SNN funding scheme.}

\dataavailability{No new data is generated in relation to this study.}

\acknowledgments{Jean Schneider is grateful to Juan Cabrera for the Figure~\ref{mutual}. We utilized Writefull's AI to help clarify certain phrasings.}

\conflictsofinterest{The authors declare no conflicts of {interest.} 
}



\begin{adjustwidth}{-\extralength}{0cm}

\printendnotes[custom]
\reftitle{References}


\PublishersNote{}
\end{adjustwidth}
\end{document}